# The stability of uniformly rotating stellar disks.

P. Vauterin* and H. Dejonghe

Universiteit Gent, Sterrenkundig Observatorium, Krijgslaan 281, B–9000 Gent, Belgium

**Abstract.** We explore a series expansion method to calculate the modes of oscillations for a variety of uniformly rotating finite disks, either with or without a dark halo. Since all models have the same potential, this survey focuses on the role of the distribution function in stability analyses. We show that the stability behaviour is greatly influenced by the structure of the unperturbed distribution, particularly by its energy dependence. In addition we find that uniformly rotating disks with a halo in general can feature spiral-like instabilities.

The beautiful spiral structure, bars and rings which disk galaxies often show point at the presence of (possibly transient) perturbations in the disk. The visual appearance probably should be explained to some extent by gas dynamics, but the stellar component must be an important player too when the spiral structure is well-developed and thus probably not too ephemeral.

The question of the stability of such disks and the structure of the instabilities which might occur is an important one, but the solution turns out to be quite involved for realistic cases. Currently, N-body calculations offer the most flexible tool for the study of the evolution of models for such galaxies. Alternatively, one models the perturbations in a galaxy by searching for linear modes, i.e. selfconsistent linear solutions of both the Poisson equation and the linearised collisionless Boltzmann equation. The results from this approach seem to be compatible with N-body simulations, at least concerning the early stage of the instability (see e.g. Sellwood & Athanassoula, 1986). Although the literature offers general methods for calculating these modes (Kalnajs, 1977), up to now the calculations only have been done for a few isolated cases.

In this article, we calculate linear modes for a particular set of simple models, i.e. flat disks with finite radius in a quadratic potential. Although these models are an oversimplification since they do not show differential rotation, they can reveal some interesting information about the stability of real galaxies, particularly in the central parts where the potential is sometimes approximately quadratic. We have chosen these models since we can apply a general and straightforward scheme to calculate the linear modes in quadratic potentials. This method is based on the series expansion approximation for the solution of the linearized Boltzmann equation (Vauterin & Dejonghe, 1995), which becomes exact in the case of quadratic potentials.

The mode analysis has already been performed analytically for uniformly rotating self-consistent disks without a halo and with some particular distributions (Kalnajs, 1972). We extend this work by calculating modes for a variety of different disk distribution functions, embedded in an inert halo.

## 1. Method

*1.1. Solutions for the linearized Boltzmann equation*

The total disk is supposed to consist of two parts, a time-independent axisymmetric unperturbed part, plus a small perturbation. The distribution function is denoted as

$$f(r, \theta, v_r, v_\theta, t) = f_0(r, v_r, v_\theta) + f'(r, \theta, v_r, v_\theta, t), \qquad (1)$$

while the total potential, which is defined as a binding energy (decreasing outwards), reads

$$V(r, \theta, t) = V_0(r) + V'(r, \theta, t). \qquad (2)$$

The unperturbed quantities are marked with a subscript 0. Moreover, using Jeans' theorem, we know that the unperturbed distribution function can also be written in terms of the two integrals of motion, the binding energy $E$ and the angular momentum $J$, defined by

$$E = V_0(r) - \frac{1}{2}(v_r^2 + v_\theta^2) \qquad (3)$$

and

$$J = r v_\theta. \qquad (4)$$

Using Poisson brackets, the linearized collisionless Boltzmann equation for small perturbations is quite compact:

$$\frac{\partial f'}{\partial t} - [f', E] = [f_0, V']. \qquad (5)$$

This fundamental equation describes how the motion of the stars in a given perturbing potential determines the evolution of the distribution function, up to the first order in the perturbation.

The right hand side of the equation can be written as

$$-\nabla_v f_0 . \nabla_r V'. \qquad (6)$$

Writing the unperturbed distribution function in integral dependence $f_0(E, J)$, this becomes

$$-\left(\frac{\partial f_0}{\partial E}(E, J)\nabla_v E + \frac{\partial f_0}{\partial J}(E, J)\nabla_v J\right).\nabla_r V', \qquad (7)$$

which can be transformed again into

$$\frac{\partial f_0}{\partial E}[E, V'] + \frac{\partial f_0}{\partial J}[J, V']. \qquad (8)$$

The operator acting on $f'$ in the left hand side of the equation (5) is linear and is "seeing" $E$ and $J$ as constants since these are integrals of the motion. Therefore we can write the solution as

$$f' = \frac{\partial f_0}{\partial E}f'_E + \frac{\partial f_0}{\partial J}f'_J, \qquad (9)$$

with $f'_E$ and $f'_J$ the solutions of the linearized Boltzmann equation having respectively $f_0(E, J) = E$ and $f_0(E, J) = J$ as unperturbed distribution function. For a given unperturbed potential $V(r)$, one only has to solve the Boltzmann equation in these two special cases and use the linear combination to calculate more general distribution functions.

In this paper, we assume a quadratic potential:

$$V_0(r) = -\frac{1}{2}\Omega_0^2 r^2, \qquad (10)$$

with $\Omega_0$ the vibration frequency of the stars in rectangular coordinates. We will use the power series expansion method (Vauterin & Dejonghe, 1995, hereafter paper I) to calculate the perturbed distribution. In this approach, the radial coordinate and the velocities are expanded in power series, while the angular dependence and the time is expanded harmonically (the circular velocities are chosen as zero points for the velocity expansion). In the linearized case, the different harmonics are independent and so the expansions can be written down for each individual harmonic $m$. Adopting the same notations as in paper I, this reads

$$f' = \sum_{l \geq 0}\sum_{j \geq 0}\sum_{k \geq 0} p^m_{l,j,k} r^k v_r^l (v_\theta - r\Omega_0)^j e^{i(\omega t - m\theta)}, \qquad (11)$$

$$f_0 = \sum_{l \geq 0}\sum_{j \geq 0}\sum_{k \geq 0} p^0_{l,j,k} r^k v_r^l (v_\theta - r\Omega_0)^j \qquad (12)$$

and

$$V' = \sum_{k \geq 0} a^m_k e^{i(\omega t - m\theta)}, \qquad (13)$$

with $\omega/m$ the pattern speed of the perturbation. When these forms are substituted in the linearized Boltzmann equation, one can collect terms with equal powers and construct a set of equations in order to determine the unknown $p^m_{l,j,k}$ (see section 2.2 of paper I). This set of equations is represented most conveniently in matrix notation, by defining the vectors

$$[P^m_{n,k}]^T = (\,p^m_{0,n,k},\quad p^m_{1,n-1,k},\quad \cdots\quad p^m_{n-1,n,k},\quad p^m_{n,0,k}\,), \qquad (14)$$

$$[\mathbf{P}^0_{n,k}]^T = (\,p^0_{0,n,k},\quad p^0_{1,n-1,k},\quad \cdots\quad p^0_{n-1,n,k},\quad p^0_{n,0,k}\,). \qquad (15)$$

In addition, we define three sets of matrices (they are defined by enumerating the nonzero elements)

$$[\mathcal{A}_n(\nu, s)]_{(n+1)\times(n+1)} : \begin{cases} A_{k,k+1} = sk \\ A_{k+1,k} = \frac{n-k+1}{s} \end{cases}, \qquad (16)$$

$$[\mathcal{B}^m_{n,p}]_{(n+1)\times n} : \begin{cases} B_{k,k} = m \\ B_{k,k+1} = ki \\ B_{k+1,k} = i(p-n+k) \end{cases} \qquad (17)$$

and

$$[\mathcal{D}^m_{n,p}]_{(n+1)\times(n+2)} : \begin{cases} D_{k,k} = m(n+2-k) \\ D_{k,k+1} = ikp \end{cases}. \qquad (18)$$

If we further rescale the perturbation pattern speed into a new parameter

$$\nu_0 = \frac{\omega - m\Omega_0}{2\Omega_0}, \qquad (19)$$

the resulting set of equations for the unknown $P^m_{n,p}$ reads in the case of a quadratic potential

$$2\Omega_0 \mathcal{A}_n(-\nu_0, i) P^m_{n,p} =$$
$$-\mathcal{B}^m_{n,p+1} P^m_{n-1,p+1} - \sum_{q=0}^{p+1} a^m_q \mathcal{D}^m_{n,q} \mathbf{P}^0_{n+1,p-q+1}. \qquad (20)$$

The matrix $\mathcal{A}_n(-\nu_0, i)$ is regular for all non–integer values of $\nu_0$. The integer values correspond to resonances. In the non–resonant case, this equation can easily be solved recursively. When $E$ and $J$ are used as unperturbed distributions (using (9), the response distributions of these simple functions are sufficient to calculate a general solution), the expansion of the unperturbed distribution contains only a few low order terms. It is then sufficient to choose a perturbing potential having a polynomial form in order to obtain a terminating expansion of the perturbed distribution. In this case, the method becomes exact. The total degree of the response is the sum of the maximum degree of the unperturbed distribution and the degree of the polynomial perturbing potential.

### 1.2. Solutions for the Poisson equation

Since we suppose that the unperturbed disk extends only up to a maximum radius $d$, it is reasonable to use the family of potential-density pairs described by Hunter (1963):

$$\rho^m_{\text{Hun},l} = \frac{1}{\pi G}\frac{g^m_l}{d}\frac{P^m_l\left(\sqrt{1-r^2/d^2}\right)}{\sqrt{1-r^2/d^2}} e^{i(\omega t - m\theta)} \qquad (21)$$

$$V^m_{\text{Hun},l} = P^m_l\left(\sqrt{1-r^2/d^2}\right) e^{i(\omega t - m\theta)}, \qquad (22)$$

with $m \geq 0$, $l \geq |m|$, $P_{l,m}$ the associated Legendre functions and

$$g^m_l = \frac{\Gamma(\frac{l+m}{2}+1)\Gamma(\frac{l-m}{2}+1)}{\Gamma(\frac{l+m+1}{2})\Gamma(\frac{l-m+1}{2})}. \qquad (23)$$

For finite disks, these densities present a complete set and can be used as a basis for the expansion of general perturbed densities.

## 1.3. Self-consistent solutions for a given harmonic m

The construction of self-consistent modes for a particular distribution and with a particular harmonic symmetry $m$ consists in the search for solutions $f'$ of both the Boltzmann and the Poisson equations.

We write a general perturbing potential as a linear combination of (22),

$$V'^m(r, \theta, t) = \sum_{p=0}^{s} a_p V_{\text{Hun}, m+2p}^m(r, \theta, t). \tag{24}$$

In principle, $s$ should be $+\infty$. In practice, one has to set a reasonable upper limit for $s$ without compromising on accuracy. This can easily be checked afterwards. Typically, $s \approx 12$ in our calculations.

Since the potentials (22) are polynomials in $r$, the series expansion method can be used to calculate the response perturbed distribution function for each of them. The result will of course depend on the pattern speed $\omega$. In addition, the integration over the velocities yields the perturbed mass density

$$V_{\text{Hun}, m+2p}^m \leadsto f'^m_{m+2p}(\omega) \leadsto \rho'^m_{m+2p}(\omega). \tag{25}$$

Each of the perturbed densities, which are also polynomials, can be expanded in the densities (21), again up to a maximum order $s$

$$\rho'^m_{m+2p} = \sum_{q=0}^{s} c^m_{q,p}(\omega) \rho^m_{\text{Hun}, m+2q} \tag{26}$$

For our models, these coefficients can be calculated analytically using the expansion formulae given by (Hunter, 1963).

Using the matrix $c_{q,p}$, one can now write the response potential resulting from the perturbation (24) as

$$\sum_{q=0}^{s} \left[ \sum_{p=0}^{s} c_{q,p}(\omega) a_p \right] V_{\text{Hun}, m+2q}^m. \tag{27}$$

For a self-consistent mode, this response is required to be equal to the original perturbing potential. This can only be achieved for particular values of the pattern rotation speed $\omega$. For general values of $\omega$, one can require the response potential only to be proportional to the perturbing one, with a scale factor $\lambda$. This results in an eigenvalue problem for the $(s+1) \times (s+1)$ response matrix $C(\omega)$, with elements $c_{q,p}(\omega)$, yielding eigenvectors $a_p$ and eigenvalues $\lambda$ which are the above mentioned scale factors. Note that in the present case, the elements of the matrix $C$ are found exactly and in a form that keeps $\omega$ analytic. However, since $C$ has an infinite dimension and is not diagonal, the eigenvalues can be calculated only approximately.

## 1.4. The eigenvalue problem for uniformly rotating potentials

One finally has to calculate the eigenvalues for all values of $\omega$ in the complex plane (we will call this relation the eigenvalue relation $\lambda(\omega)$, which is multiple-valued). Values for $\omega$ resulting in a unity eigenvalue correspond to normal modes. Generally, the search for unity eigenvalues has to be performed numerically. Normal modes with a complex $\omega$ occur in complex conjugate pairs, and one of them will have an exponentially growing amplitude, causing an unstable behaviour. On the other hand, real values for $\omega$ correspond to rotating of oscillating perturbations without any exponential growth. We call these modes stable.

In a quadratic potential, the response behaviour is particulary simple since all resonances are global: when the perturbation rotates with a pattern speed $\omega$ such that $\omega/\Omega_0$ has an integer value with the same parity as $m$, the whole disk is in resonance, while otherwise no part of the stellar distribution is resonating. This is a direct consequence of the fact that, in a harmonic potential, all stars have exactly the same vibration frequency $2\Omega_0$. So the eigenvalue relation $\lambda(\omega)$ has poles at integer values of $\omega/\Omega_0$ with the same parity as $m$ (see fig. (1) for an example).

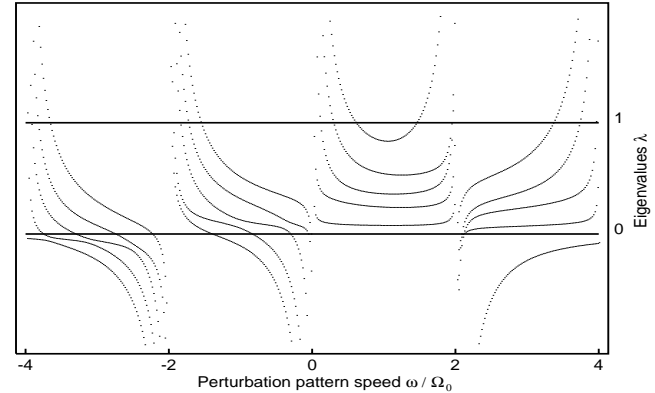

**Fig. 1.** Eigenvalue relation $\lambda(\omega)$ along the real axis for model I with $\Omega = 0.5$, $\alpha = 1$ and $m = 2$. Each curve corresponds to a branch of the characteristic equation that yields the eigenvalues.

In addition, the functional dependence of the elements of the response matrix $c_{q,p}(\omega)$ is particulary simple, since they consist only of a sum of fixed poles:

$$c_{q,p}(\omega) = \sum_{l;2} \frac{A_{q,p}}{\omega - \Omega_0 l}, \tag{28}$$

with a summation index $l$ having the same parity as $m$. This is easily shown in the framework of fourier expansions along the orbit (e.g. Kalnajs, 1977), since all orbits have one single radial period $2\Omega_0$.

The search for unity eigenvalues (corresponding to a normal mode) is equivalent to solving the equation

$$|C(\omega) - I| = 0, \tag{29}$$

where $I$ is the $(s+1) \times (s+1)$ unity matrix and $|...|$ stands for the determinant. When this equation is multiplied enough times with factors $\omega - \Omega_0 l$ to remove all poles, it becomes a polynomial function in $\omega$ with real coefficients (note that these factors can never become zero for physical responses since this would correspond to singular resonances). The degree of this polynomial is constant for a certain value of $s$ (apart from accidental cancelation of some terms, which exceptionally occurs). For distributions which have symmetrical $J$ dependence with respect to $J = 0$, this degree equals $s(s+1)/2 + sm$, while for other ones the polynomial has $s$ more roots.

The modal frequencies can now easily be found as the roots of a polynomial. This is a big computational advantage, as it is not longer necessary to perform a brute-force numerical search on the eigenvalue relation over the whole complex plane. Apart from this, it is sufficient to observe $\lambda(\omega)$ along the real axis to have a complete view of the stability behaviour. The degree of the polynomial gives the number of modes. Every pair of modes which is "missing" on the real axis should occur in the complex plane, with opposite signs for the imaginary values.

Between two poles, there are three possible situations concerning $\lambda(\omega)$ (this relation is actually multiple-valued, but holds for each branch).

- The $\lambda(\omega)$ approaches the asymptotes with opposite sign. In this case, there will always be a real mode inbetween both asymptotes.
- The $\lambda(\omega)$ approaches both asymptotes with negative sign. There can be a pair of real modes if $\lambda(\omega)$ reaches a value greater that one. We never observed this though: in all our models $\lambda(\omega)$ remained negative between such poles. [1]
- The $\lambda(\omega)$ approaches both asymptotes with positive sign. In this case the curve $\lambda(\omega)$ may or may not intersect the $\lambda = 1$ line. The corresponding frequencies are real resp. complex, and the mode is stable resp. unstable, depending on the details of the unperturbed distribution function.

In the first two cases, all modes are stable. In the last case however, unstabilities can occur for particular unperturbed distributions. This is why we will call this behaviour "potentially unstable".

## 2. The models

In the following paragraphs, we will model a galaxy using two components:

- A disk, which can be subject to perturbations. This disk has a central mass density $I_0$ and extends up to a radius $d$. The distribution function is written as $f_0(E, J)$, while the mass density is represented by $\rho_0(r)\delta(z)$ (with $\delta$ the Dirac function). This disk creates a potential which is denoted by $V_{D,0}(r)$. In general, this potential is not quadratic.

- A spherical halo, which is supposed to be unresponsive to perturbations in the disk. We represent the mass density of this halo by $\rho_H(r)$, while the potential is denoted as $V_{H,0}(r)$.

The total potential is quadratic

$$V_0(r) = V_{D,0}(r) + V_{H,0}(r) = -\frac{1}{2}\Omega_0 r^2. \tag{30}$$

In practice, we choose a particular $f_0$, defining a disk mass density $\rho_0(r)$ by integration over all velocities. This mass density produces a potential $V_{D,0}$ which is calculated using the set of potential-density pairs (21, 22) (only $m = 0$ is used). Then the spherical halo density $\rho_H$ is adjusted in order to produce an overall potential $V_0$ of the form (30):

$$\rho_H(r) = \frac{1}{4\pi G}\left[3\Omega_0^2 - \frac{1}{r^2}\frac{d}{dr}\left(r^2\frac{dV_{D,0}}{dr}\right)\right], \tag{31}$$

which expresses the spherical halo mass density for all values of $r$ between 0 and $d$. Further away the halo mass density can be chosen freely. Of course, $\rho_H$ should not be negative for any value of $r$ between 0 and $d$. For a given $V_{D,0}$ and $\Omega_0$, this yields an upper limit for the disk mass density. The halo strength will be quantified by calculating the ratio of the total mass in the halo in a sphere with radius $d$ to the total disk mass:

$$H/D = \frac{M_H(d)}{M_D} = \frac{\frac{G}{\pi}\left(3\Omega_0 d - \frac{dV_{D,0}}{dr}(d)\right)}{2\pi\int_0^d \rho_0(r)r\,dr}. \tag{32}$$

### 2.1. Rotating isotropic models with decreasing $f(E_\Omega^*)$

The algorithm proposed in the previous paragraph calculates the self-consistent modes for the disk having a general unperturbed distribution function $f_{D,0}(E, J)$. Since we want to discuss how the stability of the models depends on a few important parameters of the distribution function, we will start with a further simplification by focusing on functions which are of a form similar to those used by (Kalnajs, 1972). In order to simplify the notations, we define

$$E_\Omega^* = \frac{1}{2}(\Omega_0^2 - \Omega^2)d^2 + E + \Omega J, \tag{33}$$

with $\Omega$ a constant parameter which lies between $-\Omega_0$ and $\Omega_0$. This can be written as

$$E_\Omega^* = \frac{1}{2}(\Omega_0^2 - \Omega^2)(d^2 - r^2) - \frac{1}{2}[(v_\theta - \Omega r)^2 + v_r^2]. \tag{34}$$

In the following sections, we will set the unit of distance equal to $d$, so that the disk extends up to radius 1. We will now use distribution functions of the form

$$f_{D,0}(E, J) = \begin{cases} \sum_{i=1}^g \beta_i E_\Omega^{*\,\alpha_i} & E_\Omega^* > 0 \\ 0 & E_\Omega^* \leq 0, \end{cases} \tag{35}$$

with real powers $\alpha_i \geq -1/2$. These models are isotropic in a frame rotating with a speed $\Omega$ (note that outer edge of the distribution is not given by the escape velocity, but by $E_\Omega^* = 0$, which is a circle centered at $v_r = 0$, $v_\theta = \Omega r$). For this reason we call them "rotating isotropic". The distribution has a discontinuity at its edge if there occur terms with $\alpha_i \leq 0$. Special attention should be paid to distributions which have a discontinuity in their functional dependence, since this gives a Dirac $\delta$-function singularity in the response distribution. This singularity is integrable though, and adds only a finite contribution to the response mass density.

The Kalnajs disks are self-consistent models with a quadratic potential and with a distribution which is isotropic in a rotating frame. They are one of the standard models concerning stability of disks in literature since all results are analytical (Kalnajs, 1972). However, they have an integrable, but very unphysical, singularity at the edges. In addition, they have an "inverted" distribution function, i.e. the number of stars is decreasing for increasing binding energy. In a real galaxy, this will usually be the other way round. If the stability analysis of the spherical case is of any guidance here, one might expect that the slope of the energy dependence is an important factor for stability (Fridman & Polyachenko, 1984).

For this reason, we analysed some models for which the number of stars is everywhere a non-decreasing function of the binding energy. We have chosen a very simple set of models, having

$$df_{D,0} = \beta_1[E_\Omega^*]^\alpha, \tag{36}$$

with $\alpha$ positive and integer. The mass density is given by

$$\rho = \frac{2\pi\beta_1}{\alpha + 1}\left[\frac{1}{2}(\Omega_0^2 - \Omega^2)(1 - r^2)\right]^{\alpha+1}, \tag{37}$$

---
[1] It follows from the sequel that such modes would cause a disk to become unstable with increasing halo mass.

$$\sigma = \frac{\beta_1}{\alpha+2} \sqrt{\frac{1}{2}(\Omega_0^2 - \Omega^2)(1-r^2)}, \tag{38}$$

and the streaming velocity reads as

$$\langle v_\theta \rangle = \Omega r. \tag{39}$$

This velocity dispersion is proportional to $\Omega_0^2 - \Omega^2$. For a constant $\Omega_0$, increasing $\Omega$ means increasing streaming velocity and decreasing velocity dispersion. For $\Omega = \Omega_0$, the disk has no random motion and is supported by rotation only. In the next sections, we will call the distributions given by (36) "Model I". These disks usually require a substantial amount of halo mass in order to produce a quadratic potential. Even for the case closest to self–consistency, $\alpha = 0$, $H/D$ has a minimum value of 2.25, which means that the inert spherical halo is more than twice as heavy as the disk.

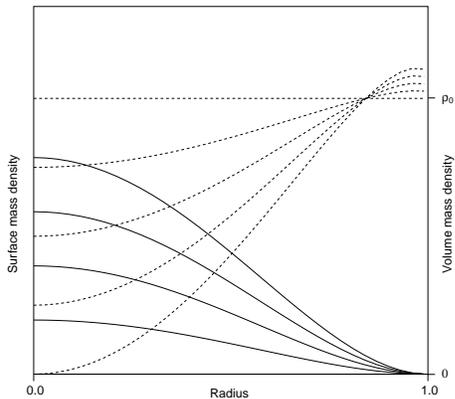

**Fig. 2.** Disk (full lines, as surface density) and halo (dashed lines, as volume density) mass density with $\alpha = 1$ for different values of $H/D$. The curves with highest disk central masses correspond to the curves with lowest halo central masses. If no disk is present, the halo has a constant density $\rho_0 = 3\Omega_0^2/4\pi G$.

In addition, we can adjust the halo/disk proportion $H/D$. Fig. 2 shows the mass density for disk and halo for different values of $H/D$ in the $\alpha = 1$ case. Note that the amount of inert material increases in the outer regions. The lowest value which is consistent with an everywhere non-negative halo density is $H/D = 6.645$. This value is so high since we have chosen a spherical halo. For an oblate or even a disky halo, this minimal value is much lower. The common intersection point for all halo mass density curves corresponds to the point where

$$r^2 \frac{dV_{D,0}}{dr}(r) \tag{40}$$

reaches a maximum.

The real axis cut of $\lambda(\omega)$ for $\alpha = 1$, $\Omega = 0.5$ and $m = 2$ is shown in Fig. 1 in a minimal halo configuration. In this case, there are no unstable modes, but there exist pairs of potentially unstable modes between $\omega = 0$ and $\omega = 2$. The qualitative behaviour of $\lambda(\omega)$ is quite representative for other cases of Model I with different $\alpha$, $\Omega$ or $m$. There is always one single region between two adjacent poles where all models have pairs of potentially unstable modes. These potentially unstable modes all occur at values of $\omega$ having the same sign as $\Omega$ (for $m = 0$ they occur at $\omega = 0$), in other words they are corotating with the streaming velocity of the galaxy.

The minimum reached by the eigenvalues in this potentially unstable region is increasing for increasing $\Omega$ and for decreasing halo-to-disk fraction $H/D$. If this minimum is greater than 1, the pair of real solutions will become a pair of complex conjugate solutions, corresponding with instabilities. For $\alpha = 0$ and $m = 2$, fig. 3 shows the positions of the stable and unstable modes for varying $\Omega$ in the minimal halo configuration. Although the $\alpha = 0$ case is a somewhat unphysical limiting case, we will present most of the results for this model since the instabilities appear at the lowest values for $\Omega$, resulting in clearer graphs. However, it is important to note that models with higher $\alpha$ show the same behaviour. Even combinations of terms with different $\alpha$ show similar behaviour as long as the distribution is an increasing function of the binding energy.

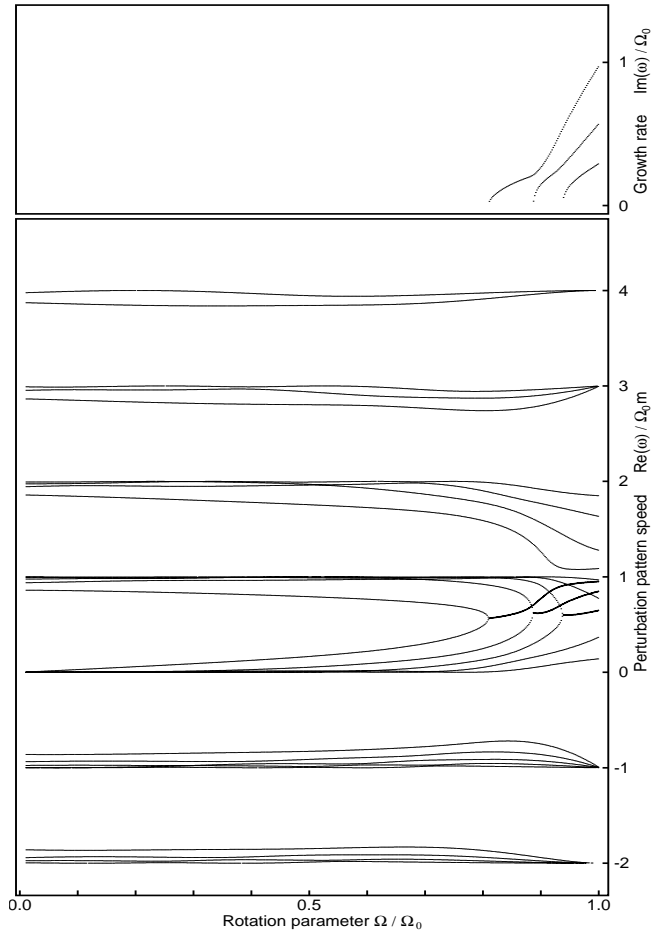

**Fig. 3.** Mode frequency diagram for Model I with $\alpha = 0$ for $m = 2$

### 2.1.1. Stable modes

For the $\alpha = 1$ configuration, Fig. 4 shows a few density profiles of the lowest order oscillating (stable) modes, resulting from the normal mode calculations. For each value of the symmetry number $m$, multiple modes are found. These modes are further catalogued using a second parameter $n$, starting at 0. For increasing $n$, the number of zero-points in the interval $[0, d]$ is increasing. Each mode $(m, n)$ has zero points interleaving the zero points does not occur. This is reasonable since the total mass contained in each mode should vanish (the disk cannot lose or gain mass). For $m = 0$, it follows immediately that the radial dependence should have at least one zero point. In general, each mode $(m,n)$ appears to occur at $2n+2$ values for $\omega$. For small values of $r$, every mode $(m,n)$ behaves like $r^m$ (see also paper I).

Only the lowest order modes have an easy physical interpretation. The $(0,1)$ mode is a so-called "breathing mode", indicating that the galaxy is periodically expanding and compressing. In the $(1,0)$ mode, the centre of the galaxy is displaced and rotates around the centre of the unperturbed potential. The $(2,0)$ mode essentially produces a rotating bar. When no halo is present, the $(1,0)$ mode does not rotate and represents a simple displacement of the system.

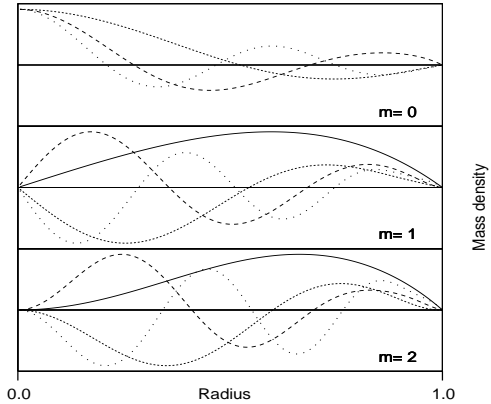

**Fig. 4.** Density profiles of the modes with lowest orders for Model I with $\alpha = 1$

### 2.1.2. Unstable modes

When $\Omega$ is large enough and $H/D$ is low enough, i.e. when the disk is cool and heavy, some of the oscillating modes become unstable. Fig. 5 shows the instability limits for the lowest order modes as a function of the rotation parameter $\Omega$ and the halo-to-disk proportion. The lower right corner contains the unstable configurations. This corresponds to cold, rapidly rotating disks and light halo's. The first instability appears at $\Omega/\Omega_0 = 0.81$. At this point, the value of Toomre's stability criterion in the centre of the disk is $Q = 1.88$, lying quite close to the value 1.68 in the case of the uniformly rotating sheet (Toomre, 1964).

Fig. 5 further shows that the stability increases rapidly with increasing $n$. For $n > 3$, there are no unstable modes found. However, for unstable modes, the number $n$ loses some of its meaning since it does not count the number of zero points in the radial density profile anymore. In fact, for values of $\Omega$ very close to $\Omega_0$, even an $n = 0$ mode contains multiple zero points, resulting in a somewhat patchy perturbation. The second parameter $n$ is therefore only defined as a continuation of the stable case.

From fig. 5, it is clear that the slopes of the instability lines are increasing as $m$ increases. As a result, for large values of $\Omega$ (meaning cooler disks), the disk becomes more unstable for larger $m$. This feature, combined with the remarks in the pre-vious paragraphs, indicate that, for relatively low values of $\Omega$ (hot disks), low order global modes ($m = 0, 1, 2$) dominate the instability space. Contrarily, for high values of $\Omega$ (cool disks), short ranged local instabilities dominate (cfr. the Jeans instability).

As can be seen from fig. 3, the pattern speed of the instabilities increases as they become more and more unstable (higher $\Omega$), but it remains smaller than the rotation speed of the stars.

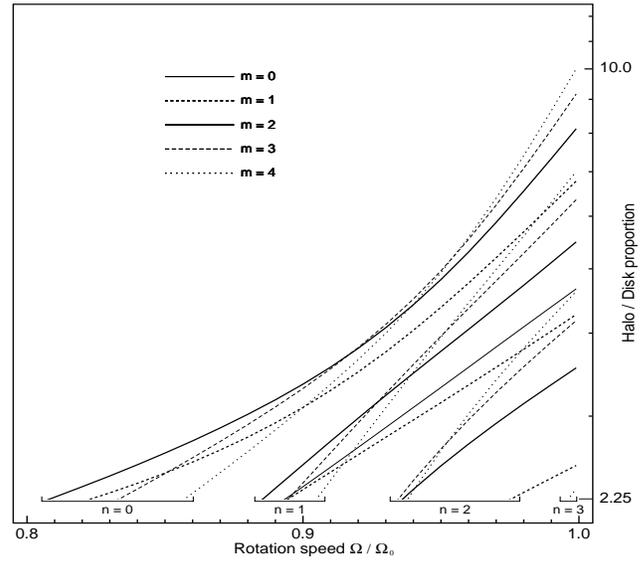

**Fig. 5.** Instability limits for Model I, $\alpha = 0$ (lower right corner= unstable region). Note that the $n = 0$, $m = 0$ modes do not occur, as explained in the text. For $n = 2$, the $m = 0$ mode is already completely stable for all $\Omega$.

Fig. 6 shows the (2,0) mode, which is the most unstable one for intermediate values of $\Omega$ and shows a spiral-like behaviour. Most of the unstable modes show a spiral-like structure and they are always trailing.

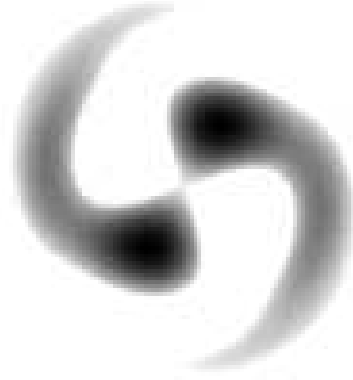

**Fig. 6.** Mass density of the unstable (2,0) model I for $\alpha = 0$ and $\Omega/\Omega_0 = 0.85$. The streaming velocity goes counterclockwise.

The presence of spiral modes is usually associated with differential rotation and the fact that the perturbation is winded in the neighbourhood of resonances. It is clear that this effect cannot explain the spiral structures in the present models, since these have no differential rotation. However, the distribution function itself can do the job since it is able to put, depending on the radius, varying weights on the various resonances. Suppose that the disk is perturbed by a rotating quadrupole potential which is aligned to the $X$-axis of the rotating frame:

$$V' = \text{Re}\left[V'(r)e^{i(2\theta-(\omega_r+\omega_c)t)}\right], \tag{41}$$

with $V'(r)$ a real function. The perturbed mass density is given as a sum of contributions from different resonances, each having the form of a pole (see also eq. 28)

$$\rho'_l = \text{Re}\left[\frac{A_l}{\omega_r + i\omega_c - 2l\Omega_0}\right]e^{2\theta i}. \tag{42}$$

If $\omega_c$ is not zero, the contribution coming from such a resonance is not aligned to the $X$-axis anymore:

$$\rho'_l = \frac{A_l}{(\omega_r - 2l\Omega_0)^2 + \omega_c^2}[(\omega_r - 2l\Omega_0)\cos 2\theta + \omega_c \sin 2\theta], \tag{43}$$

resulting in a response which lags behind the perturbing potential by an angle $\alpha$ given by

$$\tan 2\alpha = \frac{\omega_c}{2l\Omega_0 - \omega_r}. \tag{44}$$

From fig. 3, it is clear that all instabilities occur slightly below corotation (for the whole disk). This means that the lag angle is large (and positive) for CR and much lower for ILR, OLR and higher order resonances (as $2l\Omega_0 - \omega_r$ is smallest for CR).

In addition, the velocity dispersion is decreasing with increasing radius, as in most cases. This means that the outer regions are biased in favour of circular orbits, supporting mainly corotation resonance, while the central part has more eccentric orbits, giving more contribution to higher order resonances. Therefore, it can be inferred that the lag angle is increasing as $r$ increases, resulting in a trailing spiral structure.

Of course, this scheme does not conflict with the existing explanation for the formation of trailing spirals, but it rather presents an extra effect which is presumably only dominant in regions with almost no differential rotation.

### 2.2. Rotating isotropic models with increasing $f(E_\Omega^*)$

As mentioned before, the mode analysis for the Kalnajs disks was already performed fully analytically earlier (Kalnajs, 1972). These disk have an unperturbed distribution function

$$f_0 = E_\Omega^{*\,-1/2}, \tag{45}$$

In this special case (and for all weighted integrals of (45) over $\Omega$), the mode analysis is particulary simple since the response matrix $C(\omega)$ turns out to be already diagonal when the potential-density pairs (22, 21) are applied. This simplifies the calculations considerably, but puts severe limitations on the structure of the perturbations. All density profiles are real functions of the radius, excluding spiral structure.

The distribution function for the Kalnajs disks becomes infinite at its outer edges. This singularity does not cause fundamental problems in the mode analysis, but one still can ask to what extent the particular behaviour of these disks is determined by this feature. In order to address this question, we used models of the form (33), which are obviously everywhere finite, fitted to (45) and calculated the mode spectrum. Even for a relatively low value of $g$, the results are even quantitatively very similar to those of the distribution (45). Both the mode frequencies and their spatial structure are in very good correspondence. As a sideproduct, this provides a convincing check for the validity of our calculations.

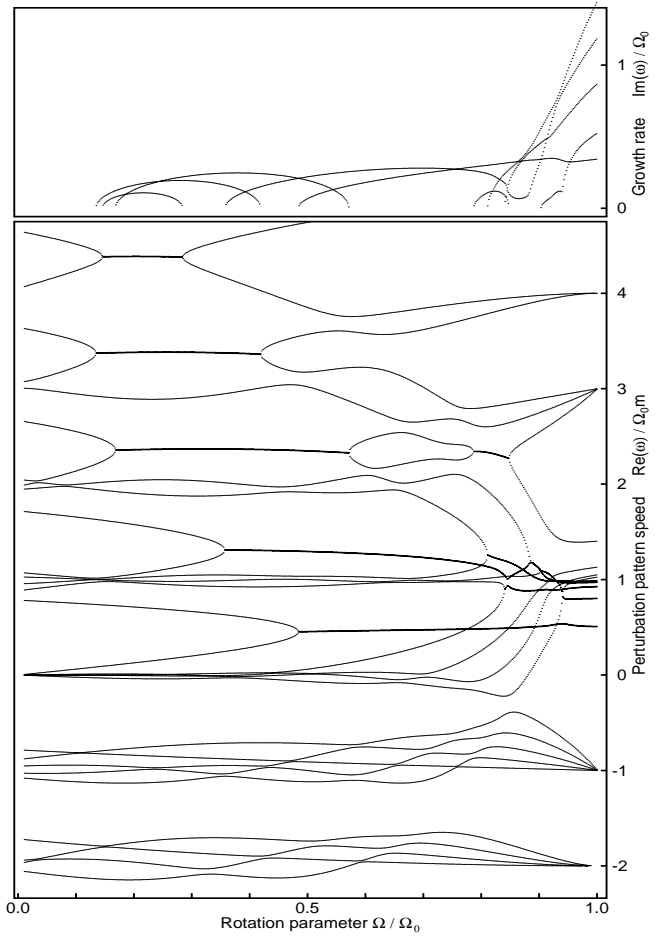

**Fig. 7.** Mode frequencies for model II with $g = 3$ and $m = 2$

Fig 7 shows the mode frequencies as a function of the rotational parameter $\Omega$. In this case, the situation is much more complicated than in the simple diagram shown in fig. 3. It is not obvious anymore that the disk becomes more unstable for increasing $\Omega$, since there exist instabilities over the whole range of $\Omega$. In the case $m = 0$, there even exist unstable modes for $\Omega = 0$. Moreover, a particular unstable mode can become stable again when $\Omega$ is increased. When compared to fig. 7, the only striking common feature are the very rapidly growing unstable modes occurring at high values for $\Omega$, which are actually the only ones that are present in Model I. This seems to indicate that these instabilities are a kind of "common" modes, in principle occurring in every uniformly rotating disk, regardless the structure of the unperturbed distribution. It is interesting to note that these "common" modes all appear close to corotation, having pattern speeds around $\Omega_0$, while most of the others have higher pattern speeds. In all the models which we analysed, these "common" modes were always present, while not a increasing function of the binding energy. The "common" modes all appear at approximately the same value for $Q$, around 1.8.

The response matrix $C$ becomes more diagonal-like as the number of components used to fit (45) increases. Correspondingly, the imaginary part of the density profiles of unstable modes becomes smaller and smaller (in the limit of the Kalnajs models, the imaginary part vanishes), making the profile more bar-like and more aligned to the perturbing potential.

### 2.3. Two-stream models

In plasma physics, it is well known that two merged components flowing with different mean velocities can cause instabilities (the so called two-stream instability). One might expect that a similar effect can occur in a galaxy consisting of multiple components with different streaming velocities (see e.g. Araki, 1987). Apart from the single-streaming Model I, we studied the stability of a galaxy resulting from the superposition of two models I with $\alpha = 1$ with equal but opposite rotation speeds

$$df_{D,0} = \frac{\beta_1}{2}(E^*_\Omega + E^*_{-\Omega}). \qquad (46)$$

These distributions give rise to exactly the same mass density as in fig. 2.

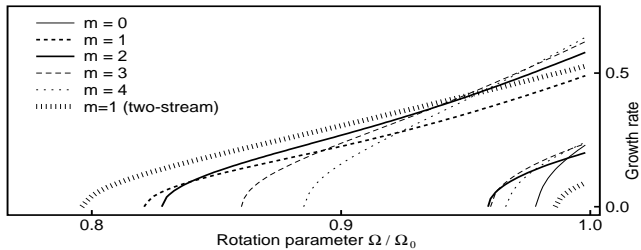

**Fig. 8.** Growth rate diagramme for Model I and two-stream Model I with $\alpha = 1$

When compared to the single-stream models, all modes are more stable except the lop-sided modes with $m = 1$ (the $m = 0$ modes are not affected), which become less stable. This is in agreement with other investigations, both linear calculations (Araki, 1987) and numerical simulations (Zang & Hohl, 1978). This $m = 1$ unstable modes occur at $Re(\omega) = 0$. Fig. 8 shows the stability diagramme for the $m = 1$ instabilities in the two-stream case, superimposed on that of the single component model.

In the case of (46), the two counterrotating components are perfectly balanced and there is no net rotation. If one of the streams is weaker than the other, the $m = 1$ mode has some remaining pattern speed and shows a one-armed spiral shape, as shown in fig. 9.

### 3. Conclusions

We have explored the stability of stellar disks embedded in a quadratic potential. The main goal is to investigate the influence of the distribution function on the stability. This has been achieved by studying the mode spectrum as a function of important parameters, such as slope of the binding energy dependence, velocity dispersion, mean rotation and disk/halo fraction.

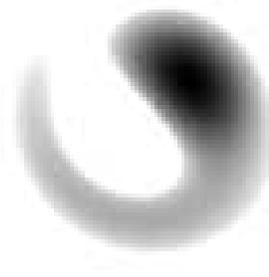

**Fig. 9.** Mass density of the lop-sided $m = 1$ instability in a partially counterrotating model I with $\alpha = 0$ and $\Omega/\Omega_0 = 0.85$. The net streaming velocity goes counterclockwise, i.e. the spiral structure is trailing.

We have demonstrated that uniformly rotating disks are by no means the boring structures which they are often taken to be: they display a very varied stability behaviour, ranging from totally stable to violently unstable disks. Primarily responsible for this is the underlying distribution function.

In our analysis, all isotropic non-rotating disks were stable, if they have a distribution function which is increasing with increasing binding energy. The Kalnajs disks prove that the opposite is certainly not true. Although this does not prove any general behaviour, it might be that, at least for uniformly rotating disks, there is some law similar to Antonov's for the spherical isotropic case (Fridman & Polyachenko, 1984). As a general behaviour, non-inverted distributions become less stable as they have more rotation velocity and less random velocity. They are stabilized again by increasing the amount of inert halo.

An other remarkable property is the occurrence of spiral mode instabilities in most of the models. Usually, the formation of spiral modes is associated with the presence of differential rotation. It may well be that these modes can be seen as so-called "edge modes" (Toomre, 1981) or "groove modes" (Sellwood & Kahn, 1991), where amplification is triggered by the presence of a relatively steep gradient in mass density in the outer regions of the disk. As explained in section 2.1.2, the absence of spiral-like modes in the Kalnajs disks seems to be a mathematical coincidence rather than a general property of uniformly rotating disks. There are big qualitative and quantitative differences in mode frequency diagrammes between different models, but they all tend to share a set of very rapidly growing modes if the disks are sufficiently cold. These common instabilities seem to constitute the set of modes which are largely independent of the structure of the unperturbed distribution.

The two-streaming version of model I is well suited to examine the consequenses of the presence of two merged counterrotating components on the stability. Of course, the stability analysis shows fundamental differences compared to the single-streaming galaxy, particularly in the shape of the responses and the pattern speeds. We confirm that the $m = 2$ and higher order modes are all considerably stabilized by the presence of the second stream. The $m = 0$ is not affected, but the $m = 1$ mode is less stable than in the single model.

*Acknowledgements.* P. Vauterin acknowledges support of the Nationaal Fonds voor Wetenschappelijk Onderzoek (Belgium)

## References

Araki, S., 1987, Astron. J., 94, 99
Fridman, A. M. & Polyachenko, V. L., 1984, Physics of Gravitating Systems, Springer-Verlag.
Hunter, C., 1963, MNRAS, 126, 299
Kalnajs, A. J., 1972, ApJ, 175, 63
Kalnajs, A. J., 1977, ApJ, 212, 637
Sellwood, J. A., Athanassoula, E., 1986, MNRAS, 221, 195
Sellwood, J. A., Kahn, F. D., 1991, MNRAS, 250, 278
Toomre, A., 1964, ApJ, 139, 1217
Toomre, A., 1981, In Structure and evolution of normal galaxies, eds. S. M. Fall & D. Lynden-Bell, Cambridge University press, 111
Vauterin, P. & Dejonghe, H., 1995, A & A, in press (paper I)
Zang, T. A. & Hohl, F., 1978, ApJ, 226, 521